\documentstyle[floats,aps]{revtex} 
\begin{document} 
\input epsf 
 
  \font\twelvemib=cmmib10 scaled 1200 
  \font\elevenmib=cmmib10 scaled 1095 
  \font\tenmib=cmmib10 
  \font\eightmib=cmmib10 scaled 800 
  \font\sixmib=cmmib10 scaled 667 
  \skewchar\elevenmib='177 
  \newfam\mibfam 
  \def\mib{\fam\mibfam\tenmib} 
  \textfont\mibfam=\tenmib 
  \scriptfont\mibfam=\eightmib 
  \scriptscriptfont\mibfam=\sixmib 
\def\frac#1#2{{\textstyle{#1 \over #2}}} 
\def\pz{\partial} 
\def\xhi{{\raise.35ex\hbox{$\chi$}}} 
\def\nd{^{\vphantom{\dagger}}} 
\def\ns{^{\vphantom{*}}} 
\def\yd{^\dagger} 
\def\undertext#1{$\underline{\hbox{#1}}$} 
\def\ie{{\it i.e.\/}} 
\def\etal{{\it et al.\/}} 
\def\eg{{\it e.g.\/}} 
\def\vth{\vartheta} 
\def\half{\frac{1}{2}} 
\def\fourth{\frac{1}{4}} 
\def\Tr{\mathop{\rm Tr}} 
\def\ket#1{{\,|\,#1\,\rangle\,}} 
\def\bra#1{{\,\langle\,#1\,|\,}} 
\def\braket#1#2{{\,\langle\,#1\,|\,#2\,\rangle\,}} 
\def\expect#1#2#3{{\,\langle\,#1\,|\,#2\,|\,#3\,\rangle\,}} 
\gdef\journal#1, #2, #3, 1#4#5#6{{\sl #1~}{\bf #2}, #3 (1#4#5#6)} 
\def\pra{\journal Phys. Rev. A, } 
\def\prb{\journal Phys. Rev. B, } 
\def\prl{\journal Phys. Rev. Lett., } 
\def\Nphi{{N_\phi}} 
\def\abar{{\bar a}} 
\def\bbar{{\bar b}} 
\def\cS{{\cal S}} 
\def\cN{{\cal N}} 
\def\bfr{{\mib r}} 
\def\bfk{{\mib k}} 
\def\bfq{{\mib q}} 
\def\bfp{{\mib p}} 
\def\qhat{{\hat q}} 
\def\zhat{{\hat {\bf z}}} 
\def\rhobar{{\overline{\rho}}} 
\def\sigm{U$(N)/[$U$(m)\times$U$(N-m)$] } 
\def\rmL{{\rm L}} 
\def\rmR{{\rm R}} 
\def\uar{\uparrow} 
\def\dar{\downarrow} 
\def\klu{\ket{\rmL\uar}} 
\def\kld{\ket{\rmL\dar}} 
\def\kru{\ket{\rmR\uar}} 
\def\krd{\ket{\rmR\dar}} 
\def\eps{\epsilon} 
\def\cH{{\cal H}} 
 
\draft 
 
\twocolumn[\hsize\textwidth\columnwidth\hsize\csname 
@twocolumnfalse\endcsname 
\title{SU($N$) Quantum Hall Skyrmions} 
\author{D. P.  Arovas$^1$, A. Karlhede$^2$, and D. Lillieh\"o\"ok$^2$} 
 
\address{ 
$^1$Department of Physics, 
University of California at San Diego, 
La Jolla, CA 92093, USA 
} 
 
\address{ 
$^2$Department of Physics, 
Stockholm University, 
Box 6730, S-11385 Stockholm, 
Sweden 
}

\date{\today} 
\maketitle 
\begin{abstract} 
We have investigated skyrmions in $N$-component quantum Hall systems. We find  
that SU($N$) skyrmions are the lowest energy charged excitations for filling  
factors $\nu = 1,2,\ldots,N-1$ for small enough symmetry breaking terms.  
$N>2$ skyrmions can be realized in Si QH systems based on the (110) or (111)  
interfaces of Si, or perhaps in Si (100) systems, where 
the spin and valley isospin together provide an SU(4)-symmetry,  or in  
multilayer QH systems.  
We also present Hartree-Fock results for a phenomenological easy-axis 
SU($2$)-breaking model appropriate to valley degeneracy. 
\end{abstract} 
\vskip2pc] 
 
\narrowtext 
 
Textured quasiparticle excitations in the quantum Hall (QH) effect  
involving nontrivial configurations of electron spin (``skyrmions'') 
were proposed in \cite{sondhi1}.  There is now strong experimental evidence 
that such excitations are the lowest energy charged quasiparticles at 
filling fraction $\nu=1$ \cite{barrett1,schmeller,goldberg}. 
The physics behind these excitations is as follows.  First, even in the 
absence of Zeeman coupling, the QHE system is an exchange ferromagnet, owing 
to the degeneracy of the Landau levels.  Second, spin textures carry charge; 
the charge density is proportional to the skyrmion (topological) density.   
Finally, there is a competition between 
Coulomb energy, which wants to make skyrmions of infinite size, and the 
Zeeman energy, which wants to shrink the skyrmions to a point; this 
determines the size of the skyrmion.  (While such topological excitations 
always exist, they are not necessarily the lowest energy charged excitations 
\cite{xgwu}.)  These textures exist also in other two-component QHE systems, 
such as quantum Hall bilayers, in which the layer index plays the role of 
a pseudospin \cite{indiana}. 
 
Here we study skyrmions in multi-component QH systems 
\cite{gmreview}.  We consider the case where there are $N$ degenerate 
internal ``orbitals'' for each electron, leading to an SU($N$) symmetry. 
Our principal motivation for this generalization is Si QH systems, 
in which, in addition to spin, there is a valley degeneracy for each single 
particle state \cite{rasoltbook}.  The valley degeneracy depends on the 
orientation of the interface along which the two-dimensional electron 
gas lives.  For a (100) interface,  which is the orientation usually  
used in Si MOSFET QH systems, the six valleys are split into an 
upper quadruplet and a lower doublet.  The doublet is then further split 
by the interfacial electric field $E$, analogous to the Zeeman splitting 
of up and down spin states in a magnetic field.  However, for 
a (110) interface the quadruplet lies lower and is split 
into two doublets by $E$, and in the case of a (111) interface  
all six valleys remain degenerate.  
With $N_{\rm v}$ degenerate valleys and two 
spin states, the appropriate symmetry group, in the absence of Zeeman and 
other symmetry-breaking terms, is SU($N$) where $N=2N_{\rm v}$. 
 
Our main results are as follows.  We derive the energy functional for  
SU($N$) QH skyrmions at filling factor $\nu = 1,2,\ldots,N-1$ and show 
that the energy required to create a skyrmion-antiskyrmion pair is half 
the energy to create a (polarized) quasielectron-quasihole pair.  In  
addition to their charge, the skyrmions are characterized by $N-1$ quantum 
numbers. For Si (100) QH systems for example, this implies that for small enough 
symmetry-breaking terms SU($4$) skyrmions would be the 
lowest energy charged excitations at $\nu = 1,2,3$. However, this limit is hard 
to reach. Better candidates for SU(4) and SU(6) skyrmions are Si (110) and Si (111)  
QH systems respectively. Multilayer QH systems offers another possibility to realize  
$N>2$ skyrmions.
We consider briefly  
symmetry breaking terms and determine the energies for valley SU($2$)  
skyrmions as a function of the strength of an easy axis symmetry breaking term. 
 
{\it SU($N$) QH Skyrmions} -- We first generalize the microscopic derivation 
of the skyrmion energy and charge density of Yang \etal \cite{indiana} to 
the case of $N$ electron ``flavors''.  This will allow us to map the low-energy 
dynamics onto a nonlinear sigma model.  Let $\mu=1,\ldots,\Nphi$ index the 
degenerate lowest Landau level single particle states on the plane;  
$\Nphi$ is the total number 
of flux quanta.  Consider now the Slater determinant state 
$\ket{\Psi_m[0]}=\prod_{a=1}^m\prod_{\mu=1}^{\Nphi}\psi\yd_{\mu a}\ket{0}$ 
of integer filling factor $\nu=m\le N$.  We define the {\it first\/}-quantized 
operators $S_{\alpha\beta}(i)$ in terms of their action in a flavor-diagonal 
basis for a given electron $i$: $S_{\alpha\beta}(i)$ annihilates any state in 
which the flavor of electron $i$ is not $\beta$ and otherwise changes the 
flavor from $\beta$ to $\alpha$.  Here $i\in\{1,\ldots,\cN\}$, where 
$\cN=m\Nphi$ is the total electron number.  These operators satisfy SU($N$) 
commutation relations $[S_{\alpha\beta},S_{\rho\lambda}]=\delta_{\alpha\lambda} 
S_{\rho\beta}-\delta_{\rho\beta}S_{\alpha\lambda}$ for each electron and commute 
for different electrons.  Note that $\Psi_m$ is invariant under the U$(m)\times$ 
U$(N-m)$ subgroup generated by $\cS_{ab}$ and $\cS_{\abar\bbar}$, where 
$a,b\in\{1,\ldots,m\}$, $\abar,\bbar\in\{m+1,\ldots,N\}$, and 
$\cS_{\alpha \beta}=\sum_i S_{\alpha \beta}(i)$.  Following \cite{indiana} 
and \cite{read}, we define the flavor texture 
\begin{eqnarray} 
\ket{\Psi_m[q]} 
&=&\exp\!\int\!\! d^2\!r\left(q\ns_{a\abar}(\bfr)\, 
\overline{S_{\abar a}(\bfr)}-q^*_{a\abar}(\bfr)\, 
\overline{S_{a\abar}(\bfr)}\right)\ket{\Psi_m[0]}\nonumber\\ 
&=&\exp\!\int\!\!{d^2\!k\over(2\pi)^2}\,\sum_{i=1}^{\cN}\Big(\qhat\ns_{a\abar} 
(\bfk)\,S_{\abar a}(i)\nonumber\\ 
&&-\qhat^*_{a\abar}(-\bfk)\,S_{a\abar}(i)\Big) 
\times \rhobar_i(-\bfk)\ket{\Psi_m[0]} 
\end{eqnarray} 
where $q_{a\abar}(\bfr)$ is a complex $m\times(N-m)$ matrix-valued function 
of $\bfr$ and $\qhat_{a\abar}(\bfk)$ is its Fourier transform, 
\begin{equation} 
S_{\alpha\beta}(\bfr)=\sum_{i=1}^{\cN} S_{\alpha\beta}(i)\, 
\delta(\bfr-\bfr_i)\ , 
\end{equation} 
and $\overline{{\cal O}}$ denotes the lowest Landau level projection of 
${\cal O}$.  The projected density operator satisfies 
\begin{eqnarray} 
\overline{\exp(-i\bfk\cdot\bfr_i)}&=&\rhobar_i(\bfk) 
\nonumber\\ 
\rhobar_i(\bfk)\,\rhobar_i(\bfp)&=&\exp(\half\bfk\cdot\bfp\,\ell^2+ 
\frac{i}{2}\cdot\bfk\wedge\bfp\,\ell^2)\, 
\rhobar_i(\bfk+\bfp) 
\end{eqnarray} 
with $\ell=\sqrt{\hbar c/eB}$ the magnetic length and $\bfk\wedge\bfp 
\equiv\zhat\cdot\bfk\times\bfp$. The ground state is homogeneous, hence 
\begin{equation} 
\rhobar(\bfk)\ket{\Psi_m[0]}=\delta(\bfk)\cN\ket{\Psi_m[0]}\ , 
\end{equation} 
where $\rhobar(\bfk)=\sum_i\rhobar_i(\bfk)$ is the total projected density  
operator. 
The $q_{a\abar}$ are 
complex coordinates parametrizing the coset space U$(N)/[$U$(m)\times$U$(N-m)$], 
which is the target space of the sigma model. Since $\Pi_2$ of the target  
space is $Z$, topological excitations - skyrmions - 
characterised by {\it one} integer topological charge exist for any $N$ and $m$. For 
$m=1$ the target space is $CP^{N-1}$, 
which for $N=2$ is $S^2$. 
 
Following \cite{indiana} we calculate the excess charge density and the 
energy in the flavor texture state in a gradient expansion of $q$: 
\begin{eqnarray} 
\delta\rho(\bfr)&=&\expect{\Psi_m[q]}{\overline{\rho(\bfr)}}{\Psi_m[q]} 
- \expect{\Psi_m[0]}{\overline{\rho(\bfr)}}{\Psi_m[0]}\nonumber\\ 
&=&{i\over 2\pi}\,\nabla q_{a\abar}\wedge\nabla q^*_{a\abar} + \ldots\\ 
E&=&\expect{\Psi_m[q]}{\overline{V}}{\Psi_m[q]}- 
\expect{\Psi_m[0]}{\overline{V}}{\Psi_m[0]}\nonumber\\ 
&=&2\rho_{\rm s}^\circ\int\!d^2\!r\,|\nabla q_{a\abar}|^2 + \ldots 
\end{eqnarray} 
where 
\begin{eqnarray} 
\overline{V}&=&\half\int\!{d^2\!k\over (2\pi)^2}\,{\hat v}(\bfk)\, 
\rhobar(-\bfk)\,\rhobar(\bfk)\\ 
\rho_{\rm s}^\circ&=&-{1\over 32\pi^2}\int_0^\infty\!\!\!dk\,k^3\,h(k)\, 
{\hat v}(k)\\ 
h(k)&=&{1\over\cN}\expect{\Psi_m[0]}{\rhobar(-\bfk)\,\rhobar(\bfk)} 
{\Psi_m[0]}\ .\nonumber 
\end{eqnarray} 
Here  ${\hat v}(q)$ is the Fourier transform of the interparticle potential. 
For Coulomb interactions $\rho_{\rm s}^\circ =e^2/(16\sqrt{2\pi}\eps\ell)$.  
 
To go beyond the gradient expansion is a tedious proposition.  However, 
we may recognize the above results as those obtained from expanding 
the energy $E$ and topological density $J^0$ of the \sigm  sigma model. 
A general element in the coset space may be written as $R=U\yd\Lambda U$, 
where $U=\exp\left(\matrix{0&q\cr -q^{\dagger}&0\cr}\right)$ and  
$\Lambda=\left(\matrix{1_m&0\cr0&-1_{N-m}\cr}\right)$ ($q$ is the matrix 
$q_{a\abar}$ and and $1_p$ is the $p$-dimensional unit matrix).  The 
topological current is written 
\begin{eqnarray} 
J^\mu={i\over 16\pi}\,\eps^{\mu\nu\lambda}\,\mathop{\rm Tr} 
(R\,\pz_\nu R\,\pz_\lambda R)\ ; 
\end{eqnarray} 
the total topological charge, $Q=\int\!d^2\!r\,J^0$, is an integer. 
The energy functional is given by 
\begin{equation} 
E_0[R]=\fourth\rho_{\rm s}^\circ\int\!\! d^2\!r\,\mathop{\rm Tr}(\nabla R)^2\ . 
\end{equation} 
Our results are consistent with these, upon identifying $\delta\rho=J^0$, 
to lowest order in the gradient expansion.  As shown by \cite{indiana}, 
the long-ranged Coulomb interaction appears at higher order, and a more 
accurate energy functional is therefore 
\begin{eqnarray} 
E[R]&=&\fourth\rho_{\rm s}^\circ\int\!\! d^2\!r\,\mathop{\rm Tr}(\nabla R)^2 
\nonumber\\ 
&&\qquad+\half\int\!d^2\!r\int\!d^2\!r'\,J^0(\bfr)\,v(\bfr-\bfr')\, 
J^0(\bfr')\ . 
\end{eqnarray} 
 
The dynamics are determined by the kinetic contribution 
\begin{eqnarray} 
\expect{\Psi_m[q]}{i{\pz\over\pz t}}{\Psi_m[q]}&=& 
{i\over 2\pi\ell^2}\int\!d^2\!r\,q^*_{a\abar}\, \pz_t q_{a\abar} 
+\ldots\\ 
&\simeq&{i\over 8\pi\ell^2}\int\!d^2\!r\int_0^1\!\!\!du\, 
\mathop{\rm Tr} \left( R\,\pz_u R\,\pz_t R\right)\ ,\nonumber 
\end{eqnarray} 
where $R(u=0,t)=\Lambda$ and $R(u=1,t)\equiv R(t)$. 
The last line of the above equation is the appropriate generalization of 
the $m=1$ nonlinear sigma model kinetic term (see \cite{read}).

{\it Skyrmion and quasiparticle energies} -- We now consider the energies 
of skyrmions. 
From $\mathop{\rm Tr}|\pz_\mu R \pm i\,\eps_{\mu\nu}R\,\pz_\nu R|^2\geq 0$  
we find $E_0 \geq 4\pi \rho _s^\circ |Q|$. The minimum energy for a skyrmion  
(or antiskyrmion) with charge $Q=\pm 1$ (electric charge $\pm e$) at $\nu=m$ 
is thus $E_{\rm sk}=4\pi \rho_{\rm s}^\circ =\fourth\sqrt{\frac{\pi}{2}} 
(e^2/\eps\ell)$ (assuming Coulomb interactions). Comparing this to the energies 
of a polarized quasielectron, $E_{\rm e} = 0$, and a polarized quasihole,  
$E_{\rm h} = \sqrt{\frac{\pi}{2}}(e^2/\eps\ell)$, we see that, as in the 
SU(2) case, the gap to creating a skyrmion-antiskyrmion pair is half the gap  
to creating a pair of polarized quasiparticles.  Note that the energies are  
independent of $N$ and of $m$. 
We conclude that, for small enough symmetry breaking terms, skyrmions will 
be the lowest energy quasiparticles for $\nu = m = 1,2,3,\ldots,N-1$. 
Without the 
Coulomb term, the sigma model is scale invariant and skyrmions of any size 
with energy $E = 4\pi \rho_{\rm s}^\circ |Q|$ exist.  Coulomb interactions 
make the skyrmions infinitely large (the energy being the same).  Symmetry 
breaking terms imply a cost to texturing spins, hence they favor small 
skyrmions.  The energy is then larger than the lower bound and the size of the 
skyrmion is determined by a competition between the Coulomb interaction and 
the symmetry breaking terms.  
 
In addition to charge, the SU(2) skyrmions are characterized by their spin, 
\ie,  by their ${\cal S}^z$ eigenvalue.  For SU$(N)$ there are $N-1$ commuting 
generators, hence these skyrmions are characterized by $N-1$ quantum numbers. 
Explicit symmetry breaking terms may of course reduce the number of conserved 
quantities.

{\it Experimental Realizations} -- We have shown that SU($N$) skyrmions are the  
lowest energy  
quasiparticles in any $N$-component QH system provided the symmetry breaking terms are  
small enough.   In the ordinary Si MOSFET QH systems, based on the (100) interface, 
the combination of spin and valley degeneracy leads, 
as pointed out above, to an SU(4) symmetry.  However, it is important to note that  
the Zeeman 
term that breaks the SU(2)$_{\rm spin}$ subgroup under normal conditions  
is too large for spin skyrmions to be the lowest energy charged excitations.  
If $\tilde g= g\mu_{\scriptscriptstyle \rm B} B/(e^2/\eps\ell)>0.054$, 
then the lowest energy quasiparticles 
are fully polarized \cite{sondhi1}.  In GaAs, this criterion is not met until  
$B\approx 25$T, 
but in Si, where $g\simeq 2$, the critical field is much lower: 
$B_{\rm c}\approx 1.5$T.  Thus, under normal conditions,  
the quasiparticles are fully spin polarized throughout the quantum Hall regime and 
one may not be able to reach the SU(4) skyrmion regime.  
 
We believe the chances are best to realize  $N>2$ skyrmions in   
QH systems based on the (110) or (111) interfaces of Si. It is, for example, possible 
to grow high quality SiGe heterostructures of this type, and the symmetry breaking  
terms are believed to be small and tunable. This would give four and six 
degenerate valleys respectively and hence lead to SU(4) and SU(6) skyrmions. 
Multilayer QH systems, possibly with spin, offer another possibility to realize $N>2$  
skyrmions, if the symmetry breaking terms can be made small enough. 
 
What happens in a particular system depends crucially on the type and size of the  
symmetry breaking terms. As an example of a possible scenario we discuss in some detail  
the (100) Si case, even though it may be hard to reach the SU(4) skyrmion  
regime in such systems.  
 
{\it SU(4) and (100) Si QH Systems} -- 
Ignoring symmetry breaking terms there are  four degenerate lowest Landau 
levels and skyrmions are the lowest energy charged excitations 
at $\nu = 1,2,3$.  If we denote the valley index by left (L) and right 
(R) and spin by up ($\uar$) and down ($\dar$), a basis of  states is  
\{$\klu$, $\kld$, $\kru$, $\krd$\}, and  
the three commuting SU(4) generators can be   
choosen  as $\sigma_{zR}={\rm diag}(0,0,1,-1), \  
\sigma_{zL}={\rm diag}(1,-1,0,0)$ and $\tau_{\uparrow}={\rm diag}(-1,0,1,0)$.  
$\sigma_{zR/L}$ measure the $z$-component 
of the spin in valley $R/L$, $\tau_{\uparrow}$ is the difference in the 
number of  spin $\uparrow$ electrons in  the  $R$ and $L$ valleys. 
 
We assume the most  
important symmetry breaking terms in Si are: the spin Zeeman term,  
$H_g=\frac 1 2 g\mu_B 
B(\sigma_{zR}-\sigma_{zL})$, which affects only the   
SU(2)$_{\rm spin}$ subgroup (and breaks it to U(1)) and a term $H_w$, which 
affects only the SU(2)$_{\rm valley}$ subgroup and breaks it to a 
U(1) easy axis symmetry \cite{hrv}.  
(A possible choice is $H_w=w\sum_k(\tau_{ 
\uparrow k}+\tau _{\downarrow k})(\tau_{ 
\uparrow k+1}+\tau _{\downarrow k+1})$, $w \leq 0$ and $k$ numbers the 
orbital states in the lowest Landau level. Another possibility is 
(\ref{valleysb}) below.)  The Zeeman term causes the two spin $\downarrow$   
lowest Landau levels to have lower energy than the two spin $\uparrow$ ones. 
Since  $H_w$ provides an easy axis anisotropy, the $L/R$ valleys have the 
same energy.  These symmetry breaking terms commute with the three diagonal 
generators $\sigma_{L}, \ \sigma_{R}, \ \tau_{\uparrow}$, which thus still 
give good quantum numbers. 
 
Qualitatively, the behavior of the groundstate and the excitations as 
functions of the symmetry breaking parameters $\tilde g$ and $w$ is as  
follows.  For  
$\nu = 1$, the groundstate is two-fold degenerate: either all $L\downarrow$  
or all  $R\downarrow$ states are filled. At $\nu=2$, all $L\downarrow$ and  
$R\downarrow$ states are filled and at $\nu=3$,  $L\uparrow$ or $R\uparrow$  
are filled in addition.  In each case, for small enough $\tilde g$ and $-w$ 
skyrmions are the lowest energy charged excitations. However, for 
large enough  $\tilde g$ and/or 
$-w$ the skyrmion creation energy exceeds the energy to create polarized  
quasiparticles. Using Hartree-Fock, as for the $SU(2)$ case \cite{fertig},  
it should be possible to obtain the energies and quantum numbers for the 
skyrmions as functions of $\tilde g,w$ and in particular to determine the 
range of parameters for which the skyrmions are the lowest energy charged 
excitations.  For large $-w$, the problem reduces to the spin SU(2) case 
\cite{sondhi1}, whereas for large $\tilde g$ the problem instead reduces 
to the valley SU(2) case which we now discuss. 
 
{\it Silicon Valley SU(2)} -- If due to the largeness of $\tilde g$ the spins 
are fully polarized, valley-textured quasiparticles may be the lowest 
energy charged excitations in Si MOSFET quantum Hall systems. 
For Si (100) interfaces, the interface potential $v(z)$ leads to a splitting 
\cite{sn79} 
\begin{equation} 
\Delta E=\Big|\alpha\langle{\partial v\over\partial z}\rangle\Big| 
\end{equation} 
of the lower doublet, where $\alpha\approx 0.23\,$\AA, and 
where the average $\langle\partial v/\partial z\rangle$ is computed with 
respect to the envelope function $A(z)$, as described in ref. \cite{sn79}. 
This is analogous to the usual Zeeman splitting, and the energy gap 
as a function of $\Delta E$ has been calculated previously \cite{sondhi1}. 
Without recourse to techniques such as nuclear magnetic resonance, 
which has been used to detect spin skyrmions, it may be only through 
the energy gap, as measured in transport experiments through the activated 
behavior of $\rho_{xx}$, that valley skyrmions can be detected. 
Varying the interfacial potential $v(z)$ with an applied electric field 
or by pressure will lead to a change in $\Delta E$ and hence of the 
skyrmion-antiskyrmion creation energy. 
 
Typically, though, the Si valley splitting $\Delta E$ is quite 
small -- on the order of about 2 K -- about 1\% of the Coulomb energy 
$e^2/\eps\ell$ in Si at $B=10$.  It seems likely then that disorder-induced 
intervalley scattering will dominate this Zeeman term.  This is described 
by a Hamiltonian of the form 
\begin{equation} 
\cH'=\sum_{j=1}^{N_{\rm imp}}\left[ U\ns_j \tau^+(\bfr_j)+ 
U^*_j \tau^-(\bfr_j)\right] 
\end{equation} 
where $\tau^\pm$ are the valley raising and lowering operators. 
The disorder thus enters as a random in-plane magnetic field 
which breaks SU(2)$_{\rm valley}$ down to an easy axis  
U(1) \cite{hrv,mp}.  As we are primarily interested in the effects of 
this internal (rather than translational) symmetry breaking, we simplify 
the model by annealing the disorder in both space and (imaginary) time, 
and consider the following phenomenological Hamiltonian, 
which contains the correct symmetry breaking terms,

\begin{eqnarray} 
\label{valleysb} 
\cH&=&\half\int\!\!d^2\!r\!\!\int\!\!d^2\!r'\>V(\bfr-\bfr')\,\colon 
(\rho(\bfr)-\rho_0) 
\,(\rho(\bfr)-\rho_0)\,\colon\nonumber\\ 
&&\quad +\half\int\!\!d^2\!r\!\!\int\!\!d^2\!r'\>W(\bfr-\bfr')\, 
\colon \tau^z(\bfr)\,\tau^z(\bfr')\,\colon\ , 
\end{eqnarray} 
where 
\begin{eqnarray} 
\rho(\bfr)&=&\sum_{\mu_1,\mu_2,\alpha}\varphi^*_{\mu_1}(\bfr)\, 
\varphi\ns_{\mu_2}(\bfr) 
\,\psi\yd_{\mu_1\alpha}\psi\nd_{\mu_2\alpha}\nonumber\\ 
\tau^z(\bfr)&=&\sum_{\mu_1,\mu_2,\alpha}\varphi^*_{\mu_1}(\bfr)\, 
\varphi\ns_{\mu_2}(\bfr) 
\,\alpha\,\psi\yd_{\mu_1\alpha}\psi\nd_{\mu_2\alpha}\ . 
\end{eqnarray} 
Here we suppress the spin label (we assume full spin polarization); 
$\alpha=\pm 1$ is the valley label, and $\varphi_{\mu}(\bfr)$ is the 
normalized symmetric gauge wavefunction of angular momentum $\mu$ in the lowest 
Landau level.  We take $V(r)=e^2/\eps r$ and  
\begin{equation} 
W(\bfr-\bfr')=\left[W_0 \ell^{-2}\,\delta(\bfr-\bfr')+W_1\nabla^2 
\delta(\bfr-\bfr')\right]{e^2\over\eps\ell}\ . 
\end{equation} 
$W_0$ and $W_1$ are phenomenological symmetry-breaking parameters. 
In the easy axis case, $W_1<0$, the symmetry breaking term leads to a gap 
$-{2\over \pi} W_1 {{e^2} \over {\epsilon \ell}}$ in the dispersion relation
for the valley waves \cite{hrv}. $W_0$ is a local term that does not 
affect the form of the quasiparticles.

 
We now solve for the Hartree-Fock valley-textured quasiparticles, following 
the work of Fertig \etal \cite{fertig}.  The Hartree-Fock wave functions  
for the skyrmion and anti-skyrmion are: 
\begin{eqnarray} 
\ket{{\rm sk}}&=&\prod_{\mu=0}^\infty(u_\mu\psi\yd_{\mu+}+ 
v_\mu\psi\yd_{\mu+1,-})\psi\yd_{0-}\ket{0}\nonumber\\ 
\ket{\overline{{\rm sk}}}&=&\prod_{\mu=1}^\infty(u_\mu\psi\yd_{\mu+}+ 
v_\mu\psi\yd_{\mu-1,-})\ket{0}\ ,  
\end{eqnarray} 
where $|u_\mu|^2+|v_\mu|^2=1$ and $u_\mu,v_\mu$  are obtained by solving 
the Hartree-Fock equations self-consistently.  
Setting $u_{\mu}=1$, we obtain the  conventional quasiparticle  
and quasihole states  $\ket{\Psi_\pm}$, with energies  
\begin{eqnarray} 
E_-&=&-{W_0\over 2\pi} {e^2\over\eps\ell} \nonumber\\  
E_+&=&\left[\sqrt{\pi\over 2} -{W_1\over\pi}\right]{e^2\over\eps\ell}\ , 
\label{qpe} 
\end{eqnarray} 
relative to the energy of the polarized ground state 
$\ket{\Psi_1[0]}=\prod_{\mu=0}^\infty\psi\yd_{\mu+}\ket{0}$. 
Our Hartree-Fock results are 
plotted in figures 1 and 2. We find that valley skyrmions have lower energy 
than the polarized quasiparticles if $-0.117\simeq  W_{1{\rm c}} < W_1 \leq 0$. 
(This is true for any $W_0$ as expected since this is a local term 
\cite{indiana}.) 
 
{\it Discussion} -- To summarize,  $N$ component QH systems have an SU($N$)  
symmetry and SU($N$) skyrmions are the lowest energy charged excitations  
at $\nu=1,2,\ldots,N-1$ in the symmetric limit. $N>2$ skyrmions  
may exist as valley skyrmions in (110) and (111) Si systems, 
as combined spin and valley skyrmions in standard (100) Si systems or in  
multi-layer QH systems. The most promising candidate probably being (110)  
or (111) SiGe systems.  
 
What happens in a particular system 
depends crucially on the type and size of the symmetry breaking terms and  
experimental signatures will depend on the coupling to external probes. 
The energy gap to creating the lowest energy charged quasiparticles can be   
measured in activated transport, and for SU(2)$_{\rm spin}$ skyrmions the spin  
can be obtained from such measurements by changing the in plane magnetic field.  
An external interfacial electric field plays a similar role for  
SU(2)$_{\rm valley}$ skyrmions in Si, although in this case the energy depends 
on the field in a more complicated way. This can be used as a signal for  
SU(2)$_{\rm valley}$ skyrmions and, in combination with an in plane magnetic 
field, to  study crossovers between skyrmions with various quantum numbers 
in a limit where $\tilde g$, $-W_1$ and the valley Zeeman term are all small.

\begin{figure} [!t] 
\centering 
\leavevmode 
\epsfxsize=8cm 
\epsfysize=6.5cm 
\epsfbox{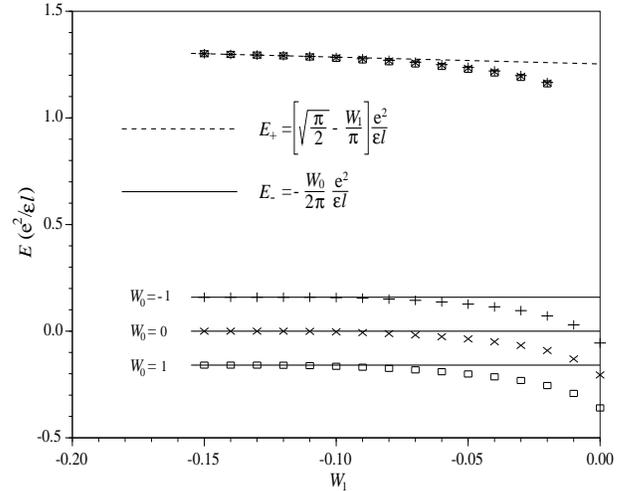} 
\caption[] 
{\label{fig1} Skyrmion (lower) and antiskyrmion (upper) energies 
{\it versus\/} $W_1$ for three values of $W_0$.  The solid and dashed 
lines correspond to the polarized quasiparticle energies of eq. \ref{qpe}.} 
\end{figure} 
 
\begin{figure} [!h] 
\centering 
\leavevmode 
\epsfxsize=8cm 
\epsfysize=8cm 
\epsfbox[18 144 592 718] {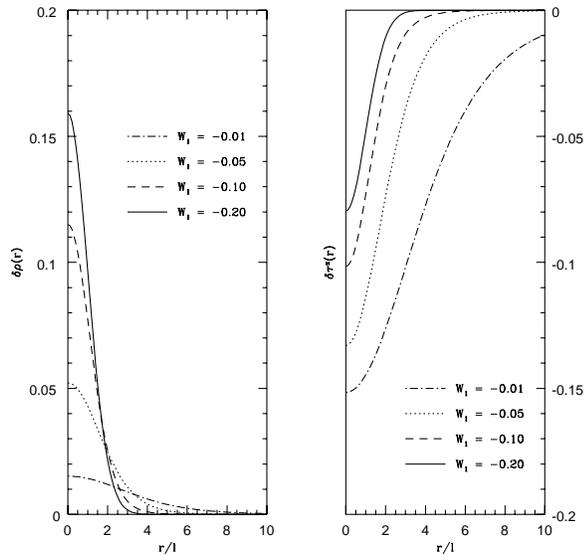} 
\caption[] 
{\label{fig2} Skyrmion particle density, $\delta \rho$, and valley density,
$\delta \tau^z$, for four values of $W_1$.}
\end{figure}

\acknowledgements 
We are grateful to S. M. Girvin, J. Furneaux, T. H. 
Hansson, K. Lejnell, A. H. MacDonald, S. Murphy, M. Shayegan and 
S. L. Sondhi for useful discussions. This work was supported in part by 
the Swedish Natural Science Research Council (AK).


\begin{references} 
 
 
\bibitem{sondhi1} S. L. Sondhi, A. Karlhede, S. A. Kivelson and E. H. Rezayi, 
{\sl Phys. Rev. B} {\bf 47}, 16419 (1993). See also, E. H. Rezayi, 
{\sl Phys. Rev. B} {\bf 36}, 5454 (1987) and {\bf 43}, 5944 (1991); 
D.-H. Lee and C. L. Kane, {\sl Phys. Rev. Lett.} {\bf 64}, 1313 (1990). 
 
\bibitem{barrett1} S. E. Barrett, G. Dabbagh, L. N. Pfeiffer, K. W. West 
and R. Tycko, {\sl Phys. Rev. Lett.} {\bf 74}, 5112 (1995); 
R. Tycko, S. E. Barrett, G. Dabbagh, L. N. Pfeiffer 
and K. W. West, {\sl Science} {\bf 268}, 1460 (1995). 
 
\bibitem{schmeller} A. Schmeller, J. P. Eisenstein, L. N. Pfeiffer and 
K. W. West, {\sl Phys. Rev. Lett.} {\bf 75}, 4290 (1995). 
 
\bibitem{goldberg} E. H. Aifer, B. B. Goldberg and D. A. Broido, 
{\sl Phys. Rev. Lett.} {\bf 76}, 680 (1996). 
 
\bibitem{xgwu} X. G. Wu and S. L. Sondhi, {\sl Phys. Rev. B} {\bf 51}, 
14725 (1995).  See also J. K. Jain and X. G. Wu, {\sl Phys. Rev. B} 
{\bf 49}, 5085 (1994). 
 
\bibitem{indiana} K. Yang, K. Moon, L. Zheng, A. H. MacDonald, S. M. Girvin, 
D. Yoshioka and S.-C. Zhang, {\sl Phys. Rev. Lett.} {\bf 72}, 732 (1994); 
K. Moon, H. Mori, K. Yang, S. M. Girvin, A. H. MacDonald, L. Zheng, 
D. Yoshioka and S.-C. Zhang, {\sl Phys. Rev. B} {\bf 51}, 5138 (1995). 
 
\bibitem{gmreview} For a recent, extensive review on multi-component  
QH systems containing references 
to the literature see, S. M. Girvin and A. H. MacDonald, ``Multi-Component 
Quantum Hall Systems: The Sum of their Parts and More'', in 
{\em Novel Quantum Liquids in Low-Dimensional Semiconductor Structures} 
edited by S. Das Sarma and A. Pinczuk (Wiley, New York, 1995). 
 
\bibitem{rasoltbook} M. Rasolt, in {\em Solid State Physics} Vol. 43, 
edited by H. Ehrenreich and D. Turnbull (Academic Press, San Diego, 1990). 
 
\bibitem{read}  N. Read and S. Sachdev, {\sl Nucl. Phys.} {\bf B316}, 609 
(1989); {\sl Phys. Rev. B} {\bf 42}, 4568 (1990). 
 
\bibitem{hrv} M. Rasolt, B. I. Halperin and D. Vanderbilt, 
{\sl Phys. Rev. Lett.} {\bf 57}, 126 (1986). 
 
\bibitem{fertig} H. A. Fertig, L. Brey, R. Cote and A. H. MacDonald, 
{\sl Phys. Rev. B} {\bf 50}, 11018 (1994). 
H. A. Fertig {\it et al.\/}, {\sl Phys. Rev. B} {\bf 55}, 10671 (1997). 
 
 
\bibitem{sn79} L. J. Sham and M. Nakayama, {\sl Phys. Rev. B} 
{\bf 20}, 734 (1979). 


\bibitem{mp} B. J. Minchau and R. A. Pelcovits, {\sl Phys. Rev. B} {\bf 32}, 
3081 (1985); A. Aharony, {\sl Phys. Rev. B} {\bf 18}, 3328 (1978). 
 
 
\end{references}
\end{document}